\documentclass[preprint,showpacs,preprintnumbers,amsmath,amssymb,prb]{revtex4}
\usepackage{graphicx}
\usepackage{dcolumn}
\usepackage{bm}
\graphicspath{{:Figs.eps:}}

\begin{document}

\preprint{APS/123-QED}

\title{Thermodynamic Properties of Model Solids with Short-ranged Potentials from Monte Carlo
Simulations and Perturbation Theory}

\author{A. D\'{\i}ez}
 \affiliation{Departamento de F\'{\i}sica Aplicada, Universidad de
Cantabria, E-39005 Santander, Spain}

\author{J. Largo}
\affiliation{Dipartimento di Fisica, Universit\`a di Roma La Sapienza, I-00185 Roma,
Italy}

\author{J. R. Solana}
\thanks{Author to whom correspondence should be addressed}
\email{ramon.solana@unican.es}
\affiliation{ Departamento de F\'{\i}sica Aplicada, Universidad de
Cantabria, E-39005 Santander, Spain}

\date{\today}

\begin{abstract}
Monte Carlo simulations have been performed to determine the excess energy and the equation
of state of fcc solids with Sutherland potentials for wide ranges of temperatures, densities and
effective potential ranges. The same quantities have been determined within a perturbative scheme by
means of two procedures: i) Monte Carlo simulations performed on the reference hard-sphere system and
ii) second order Barker-Henderson perturbation theory. The aim was twofold: on the one hand, to test
the capability of the 'exact' MC-perturbation theory of reproducing the direct MC simulations and, on
the other hand, the reliability of the Barker-Henderson perturbation theory, as compared with direct
MC simulations and MC-perturbation theory, to determine the thermodynamic properties of these solids
depending on temperature, density and potential range. We have found that the simulation data for the
excess energy obtained from the two procedures are in close agreement with each other. For the
equation of state, the results from the MC-perturbation procedure also agree well with direct MC
simulations except for very low temperatures and extremely short-ranged potentials. Regarding the
Barker-Henderson perturbation theory, we have found that, surprisingly, the first-order approximation
is in closer agreement with simulations than the second-order one.

\end{abstract}


\maketitle

\section{Introduction}

\hspace{\parindent} The Sutherland potential is of the form

\begin{equation}
u(r)=\left\{ \begin{array}{ll} \infty, & r\le \sigma\\ -\epsilon \Big(\frac{\sigma}{r} 
\Big)^{\gamma},& r>\sigma \end{array}\right.
\label{eq:Sutherland-potential} 
\end{equation}

\noindent were parameters $\sigma$ and $-\epsilon$ account for diameter of the particles and the
maximum potential depth, respectively, and the exponent $\gamma$ determines the effective
range of the  potential, so that the higher is the value of $\gamma$ the shorter is the effective
range. 

The shape of the potential function (\ref{eq:Sutherland-potential}) resembles the shape of other
widely used potentials, such as the Lennard-Jones potential, for $\gamma=6$, the Girifalco potential
for
$C_{60}$, for $\gamma\approx 12$,\cite{BS:03} and the Mie potential, widely used to describe pure
fluids and solutions as well as solids, for a wide range of values of $\gamma$ depending on the real
system to be modelled.\cite{CVK:00} The advantage of a potential of the form
(\ref{eq:Sutherland-potential}) with respect to these others is that in the former the diameter of
the particles is well defined. The  simple mathematical form of the potential
(\ref{eq:Sutherland-potential}), together with its flexibility to  reproduce a wide variety of
intermolecular interactions, has led recently to a number of applications for 
modelling interactions in complex fluids \cite{EP:02,M:04} and nanocomoposite materials. \cite{ZA:04}

In this situation, one might reasonably expect this kind of systems would have been extensively
studied bot from theory and from computer simulation. However, computer simulations on this system
have been scarcely reported. For the fluid, quite extensive simulation data for the excess energy
and the compressibility factor have been reported in ref. \cite{HW:86} and more recently in
\cite{DLS:06} for several values of $\gamma$ and a wide range of densities and temperatures,
particularly in the latter reference. Simulation data on the radial distribution function for the high
density fluid may be found in ref. \cite{DLS:07} for several temperatures and effective ranges.
Simulation data for liquid-vapour coexistence densities and critical parameters for several values
of $\gamma$ were reported in ref. \cite{CP:01} and triple point parameters in ref. \cite{C:03}.

Regarding theory, the amount of research devoted to this kind of fluids is equally scarce 
\cite{K:90,LS:00,J:05,DLS:06,DLS:07}, although recently we reported \cite{DLS:06} a quite complete
analysis of the performance of the the Barker-Henderson perturbation theory \cite{BH:67a,BH:72} for
the thermodynamic properties of this kind of fluids, and still more recently \cite{DLS:07} we
performed a similar analysis using the first-order mean spherical approximation.\cite{TL:93,TL:97}

In contrast with fluids, crystalline solids interacting by means of model potentials have been much
less frequently studied, with the possible exception of the hard-sphere solid, and to the best of our
knowledge the behaviour of solids with Sutherland potentials has not been studied before from
theory nor computer simulation. This is the aim of this paper, for which we have performed extensive
computer simulations for the thermodynamic properties of Sutherland fcc solids for several values of
$\gamma$, temperatures and densities. These are described in the next Section. Section III,
summarizes the foundations of two versions of the perturbation theory 1) an 'exact'
first-order perturbation theory based on simulations performed on the reference hard-sphere fcc solid
and the Barker-Henderson perturbation theory. These two theories are compared in Section IV with each
other and with the simulation data reported in Section II, and the concluding remarks are presented
in the same Section.

\section{Monte Carlo Simulations for fcc Solids with Sutherland Potentials}

\hspace{\parindent} We have performed {\em NVT} Monte Carlo simulations for fcc solids with
potentials of the form (\ref{eq:Sutherland-potential}) with $\gamma=6,12,18$, and $36$, reduced
temperatures
$T^{*}=0.6,0.8,1.0,1.5,2.0$ and $3.0$, and reduced densities $\rho^{*}$ in the range $0.90-1.30$ with
step $0.05$ this covering most of the density range for the solid and some of the high density
fluid. The potential cut-off distance was fixed at 
$r_{c}=3 \sigma$. Most of the systems consisted in $N=500$ particles but some simulations were carried
out with $2048$ particles to test the influence of the size of the system. The particles were placed 
in a cubic box with periodic boundary conditions in a fcc  configuration. The system was
equilibrated for $2\times 10^{4}$ cycles, each cycle consisting of an attempted move per
particle and then the thermodynamic and structural properties were measured over the next $5\times
10^{4}$ cycles, with partial averages every $500$ cycles from which the statistical errors were
determined as the standard deviation. Acceptance
ratio was fixed at around $50\%$. The excess energy $U^{E}$ was determined from the energy equation
in the form

\begin{equation}
\frac{U^{E}}{N}=2\pi\rho\int_{0}^{\infty}{g(r)r^{2}u(r)dr}
\label{eq:energy}
\end{equation}

\noindent  The compressibility factor $Z$ was determined from the expression

\begin{equation}
Z= \frac{pV}{Nk_{B}T}=1+\frac{2}{3}\pi\rho \sigma^{3}g(\sigma) +\frac{\gamma}{3}\epsilon 
\frac{U^{E}}{Nk_{B}T}
\label{eq:EOS-Sutherland-virial}
\end{equation}

\noindent whichÁh results from the combination of eq. (\ref{eq:energy}) with the
virial theorem for the Sutherland potential.

Results are listed in Tables \ref{Table:MC-gamma=6}-\ref{Table:MC-gamma=36}, in which corrections
to account for the effect of the truncation of the potential have not been included. For the
Sutherland potential with cut-off distance $r_{c}=3\sigma$, the corrections for the compressibility
factor and the excess energy are

\begin{equation}
\Delta Z=\frac{2}{3}\pi \gamma \frac{3^{3-\gamma}}{3-\gamma}\frac{\rho^{*}}{T^{*}}  
\label{eq:deltaZ-vdW-rc3}
\end{equation}

\noindent and

\begin{equation}
 \frac{\Delta U^{E}}{N\epsilon}=2\pi \frac{3^{3-\gamma}}{3-\gamma} \rho^{*} 
\label{eq:deltaUE-vdW-rc3}
\end{equation}

\noindent respectively.

\section{Perturbation Theory}

In perturbation theories, the intermolecular potential is split in the form

\begin{equation}
u(r)=u_{0}(r)+u_{1}(r),
\label{eq:u(r)-split}
\end{equation}

\noindent where $u_{0}(r)$ and $u_{1}(r)$ accounts essentially for the short-range and long-range
contributions, respectively. The former is mainly due to the repulsive forces and the latter to the
attractive ones. For the Sutherland potential (\ref{eq:Sutherland-potential}), the obvious choice for
these contributions is

\begin {equation}
u_{0}(r)=\left\{ \begin{array}{ll} \infty, & \hspace{0.5 cm} r\le \sigma\\  0, 
& \hspace{0.5 cm} r>\sigma \end{array}\right.,
\label{eq:u0(r)-S}
\end{equation}

\noindent and

\begin{equation}
u_{1}(r)=\left\{ \begin{array}{ll} 0 & \hspace{0.5 cm} r\le \sigma\\   -\epsilon 
\Big(\frac{\sigma}{r}\Big)^{\gamma},& \hspace{0.5 cm} r>\sigma \end{array}\right.,
\label{eq:u1(r)-S}
\end{equation}

\noindent respectively. Note that (\ref{eq:u0(r)-S}) is the hard sphere (HS) potential.

At high densities the thermodynamic and structural properties of a system are determined mainly by the
first of these contributions, whence arise perturbation theories. These theories
consider that the properties of the system at high densities are given by those of a reference system,
one consisting of particles interacting by means of the potential $u_{0}(r)$ and providing the main
contribution, plus a minor contribution due to $u_{1}(r)$ and considered as a perturbation of the
former. As most of the thermodynamic systems with spherically-symmetric potentials at high
temperatures behave much like a hard-sphere system, the latter is the obvious choice for the
reference system. Therefore, it is natural to express the thermodynamic properties as a series
expansion in terms of $1/T^{*}$, the inverse of the reduced temperature, with the zero-order term
corresponding to the contribution of the reference system and the remaining terms accounting for the
contribution of the perturbation. In this situation, the Helmholtz free energy, the radial
distribution function, the excess energy, and the compressibility factor can be expressed in the form

\begin{equation}
{F \over {Nk_BT}}=\sum\limits_{n=0}^\infty  {{{F_n} \over {Nk_BT}}{1 \over {T^{*^n}}}}.
\label{eq:F-per}
\end{equation}

\begin{equation}
g(r)=\sum_{n=0}^{\infty} g_{n}(r) \frac{1}{T^{*^{n}}},
\label{eq:g(r)-per}
\end{equation}  

\begin{equation}
\frac{U^{E}}{N\epsilon}=\sum_{n=0}^{\infty} \frac{U_{n}}{N\epsilon} \frac{1}{T^{*^{n}}},
\label{eq:U-per}
\end{equation}

\begin{equation}
Z=\sum_{n=0}^{\infty} Z_{n}\frac{1}{T^{*^{n}}},
\label{eq:Z-per}
\end{equation}

\noindent where subscripts '0' correspond to the contributions due to the reference 
system which, for systems with the Sutherland potential, is the hard-sphere system.

\subsection{'Exact' First-order Perturbation Theory}

Before analysing the performance of the Barker-Henderson perturbation theory for Sutherland solids, it
is worth analysing the capability of an 'exact' perturbation theory for obtaining the thermodynamic
properties of these solids. By 'exact' by mean a perturbation theory with the terms in expansions 
(\ref{eq:g(r)-per})-(\ref{eq:Z-per}) obtained from computer simulations performed on the hard-sphere
reference system, thus avoiding any theoretical approximation.

The procedure to determine the zero- and first-order terms in the expansion (\ref{eq:g(r)-per}) by
means of computer simulation was developed by Smith et al. \cite{SHB:71} quite a long time ago. They
found

\begin{equation}
g_0(r_{i}+\Delta r/2)=\frac{3\langle N_{i} \rangle_{0}}{2\pi N \rho (r_{i+1}^{3}-r_{i}^{3})}
\label{eq:g0(ri)}
\end{equation}

\noindent and

\begin{equation}
g_1(r_{i}+\Delta r/2)=-{{3\sum\nolimits_j {\left\{ {\left\langle {N_iN_j} \right\rangle _0-\left\langle {N_i}
\right\rangle _0\left\langle {N_j} \right\rangle _0} \right\}u_1^*\left( {r_j} \right)}}
\over {2\pi N\rho \left( {r_{i+1}^3-r_i^3}
\right)}}
\label{eq:g1(ri)}
\end{equation}

\noindent In the preceding expressions $N_{i}$ is the number of intermolecular distances in the range
$(r_i,r_{i+1})$, 
 with $\Delta r=r_{i+1}-r_i << \sigma$ $i=0,1,\ldots$, angular brackets mean an average, subscript
$0$ mean that the averages are  performed in the reference HS system, and
$u_{1}^{*}(r)=u_{1}(r)/\epsilon$.

Introducing the r.d.f. (\ref{eq:g(r)-per}), truncated at first order, into the energy equation
(\ref{eq:energy}) provides the first- and second-order contributions to the excess energy in the form
                                  
\begin{equation}
U_{1}=2N \pi \rho \sum_{i} g_{0}(r_{i}) r_{i}^{2} u_{1}(r_{i}) \Delta r
\label{eq:U1}
\end{equation}

\noindent and

\begin{equation}
U_{2}=2N \pi \rho \sum_{i} g_{1}(r_{i}) r_{i}^{2} u_{1}(r_{i}) \Delta r
\label{eq:U2}
\end{equation}

\noindent respectively.

Introducing in turn the excess energy (\ref{eq:U-per}), truncated at first order, into the virial
equation of sate (\ref{eq:EOS-Sutherland-virial}), provides the zero- and first-order contributions to
the compressibility factor. They are 

\begin{equation}
Z_{0}=1+\frac{2}{3}\pi\rho \sigma^{3}g_{0}(\sigma)
\label{eq:Z0}
\end{equation}

\noindent and

\begin{equation}
Z_{1}=\frac{2}{3}\pi\rho \sigma^{3}g_{1}(\sigma) +\frac{\gamma}{3}\epsilon
\frac{U_{1}}{Nk_{B}T}
\label{eq:Z1}
\end{equation}

\noindent respectively.

In principle, the procedure could be generalised to obtain higher-order terms, but the computational
effort increases with the order of the perturbative term and becomes impractical.

This procedure, that has been denoted Monte Carlo pertubation theory (MC-P),\cite{LS:04} was applied to
obtain the first terms in the perturbative expansions (\ref{eq:g(r)-per})-(\ref{eq:Z-per}) for
square-well fluids, first by Smith et al.
\cite{SHB:71}  and more recently by two of us.\cite{LS:03,LS:03a,LS:04} Very recently we have used
successfully the same procedure to obtain the thermodynamic and structural properties of fluids with
Sutherland potential.\cite{DLS:06}. To the best of our knowledge, the procedures has never been
applied to crystalline solids.

The explicit expressions of  eqs. (\ref{eq:g(r)-per})-(\ref{eq:Z-per}), truncated at the
above-mentioned level, are as follows

\begin{equation}
g\left( x \right)=g_0\left( x \right)+g_1\left( x \right){1 \over {T^*}}
\label{eq:g(x)-MC-P}
\end{equation}

\noindent where $x=r/\sigma$ is the reduced distance,

\begin{equation}
\frac{U^{E}}{N\epsilon}=\frac{U_{1}}{N\epsilon}+\frac{U_{2}}{N\epsilon} \frac{1}{T^{*}}
\label{eq:UE-MC-P}
\end{equation}

\noindent and

\begin{equation}
Z=Z_{0}+\frac{1}{T^{*}}Z_{1}
\label{eq:Z-MC-P}
\end{equation}

We have performed Monte Carlo simulations in the 
{\em NVT} ensemble of the reference HS fluid to obtain the perturbative terms in expressions
(\ref{eq:g(x)-MC-P})-(\ref{eq:Z-MC-P}). To this end, we considered a system consisting in 500 hard
spheres in an placed in a fcc configuration within a cubic box with periodic boundary conditions. At
each density, the system was allowed to equilibrate for $5\times10^{4}$ cycles and the averages
involved in expressions (\ref{eq:g0(ri)}) and (\ref{eq:g1(ri)}) were calculated from the next
$10^{6}$  cycles. We chose $\Delta r = 0.005$ and the acceptance ratio was settled to about $50 \%$.
From these data, we have calculated the values of $g_{0}(x)$ and $g_{1}(x)$ from eqs. 
(\ref{eq:g0(ri)}) and (\ref{eq:g1(ri)}), respectively, and subsequently  $U_{1}$, $U_{2}$, $Z_{0}$,
and $Z_{1}$ from eqs. (\ref{eq:U1})- (\ref{eq:Z1}). 

\subsection{Barker-Henderson perturbation theory}

The Barker-Henderson perturbation theory (BH) \cite{BH:67a,BH:72} provides expressions for the first-
and second-order terms in the expansion (\ref{eq:F-per}). The first-order term is given by

\begin{equation}
{F_{1}\over{Nk_{B}T}}=2\pi\rho\int_{0}^{\infty}{u_{1}^{*}(r) g_{0}(r)r^{2}dr}
\label{eq:F1-BH}
\end{equation}

\noindent The second-order term in the {\em macroscopic compressibility approximation}
(mc) is

\begin{equation}
{F_{2}\over{Nk_{B}T}}=-\pi\rho k_{B}T\int_{0}^{\infty}{\lbrack u_{1}^{*}(r)\rbrack^{2}
\left ({\partial\rho\over{\partial p}}\right )_{0} g_{0}(r)r^{2}dr}
\label{eq:F2-BH-mc}
\end{equation}

\noindent and in the {\em local compressibility approximation} (lc)

\begin{equation}
{F_{2}\over{Nk_{B}T}}=-\pi\rho k_{B}T\int_{0}^{\infty}{\lbrack u_{1}^{*}(r)\rbrack^{2}
\left ({\partial\lbrack \rho g_{0}(r)\rbrack\over{\partial p}}\right )_{0} r^{2}dr}
\label{eq:F2-BH-lc}
\end{equation}

\noindent The corresponding contributions to the excess energy can be
readily obtained from the thermodynamic relationship

\begin{equation}
{U^{E}\over{N\epsilon}}=-{T^{*}}^{2}\bigg\lbrack{\partial(F/Nk_{B}T)\over{\partial
T^{*}}}\bigg\rbrack_{V}
\label{eq:U-F}
\end{equation}

\noindent and those for the compressibility factor from

\begin{equation}
Z_{n}={p_{n}V\over{Nk_{B}T}}=\rho\bigg\lbrack{\partial(F_{n}/Nk_{B}T)\over{\partial\rho}}
\bigg\rbrack_{T}
\label{eq:Zn-per}
\end{equation}

Using suitable expressions for the r.d.f. $g_{0}(r)$ and the equation of state $Z_{0}$ of the HS
reference fluid, Henderson and coworkers \cite{BH:67b,BH:68,SHB:68,SHB:70,SHB:75,HSS:80}, and many
others, applied successfully the BH theory to a variety of model fluids.  

The Barker-Henderson theory has been sometimes applied to crystalline solids.
\cite{C:03,TS:97,TP:04,Z:04} To apply this theory, accurate expressions for the compressibility factor
$Z_{0}$ and the r.d.f. 
$g_{0}(r)$ of the hard-sphere reference fcc solid are needed. The equation of state of a fcc
hard-sphere solid is well reproduced by the Hall\cite{H:72} equation

\begin{eqnarray}
Z_0 & = & {3 \mathord{\left/ {\vphantom {3 \alpha }} \right. \kern-\nulldelimiterspace} \alpha
}+2.557696+0.1253077\beta +0.1762393\beta ^2-1.053308\beta ^3+2.818621\beta ^4 \nonumber\\
& & -2.921934\beta ^5+1.118413\beta ^6
\label{eq:Z-Hall}
\end{eqnarray}

\noindent where $\beta=4\alpha/(1+\alpha)$ and $\alpha=\rho_{0}/\rho-1$, in which $\rho_{0}$ is the
close-packing density.

Regarding the radial distribution function of the fcc hard-sphere solid, accurate parametrizations
are available.\cite{W:74,KW:77,RMN:96,VMN:99} One of the most frequently used is\cite{KW:77}

\begin{displaymath}
g_0\left( x \right)={A \over x}\exp \left[ {-W_1^2\left( {x-x_1} \right)^2-W_2^4\left( {x-x_1}
\right)^4}
\right]
\end{displaymath}

\begin{equation}
+{W \over {24\eta \sqrt \pi }}\sum\limits_{i=2}^\infty  {{{n_i} \over {x_ix}}\exp \left[ {-W^2\left(
{x-x_i} \right)^2}
\right]}
\label{eq:g0(x)}
\end{equation}

\noindent where $\eta=(\pi/6)\rho^{*}$ is the packing fraction and $x_{1}$ is the position of the
first peak of the r.d.f., which was determined\cite{W:74,KW:77} from the simulation data for several
packing fractions. Functions $W_{1}$, $W_{2}$, and $W$ are given by\cite{KRR:86}\\

\noindent For $\eta \le 0.55$:

\begin{equation}
W_1=\sqrt 3W_2={{\sqrt 3} \over {0.50552}}\exp \left[ {10.49375\left( {\eta -0.52} \right)}
\right]
\label{eq:W1-1}
\end{equation}

\begin{equation}
W={{{{\sqrt 3} \mathord{\left/ {\vphantom {{\sqrt 3} 2}} \right. \kern-\nulldelimiterspace} 2}}
\over {\left[ {0.23601\left( {\eta -\eta _0} \right)^2-21.4395\left( {\eta -\eta _0} \right)^5}
\right]^{{1 \mathord{\left/ {\vphantom {1 2}} \right. \kern-\nulldelimiterspace} 2}}}}
\label{eq:W-1}
\end{equation}

\noindent \\For $0.55<\eta\le 0.73$:

\begin{equation}
W_1={{1.5522782} \over {\eta _0-\eta }}-2.0302556\exp \left[ {5.8331273\left( {\eta _0-\eta }
\right)} \right]+74.873192\left( {\eta _0-\eta }
\right)^2
\label{eq:W1-2}
\end{equation}

\begin{equation}
W_2={{0.9559565-5.855022\left( {\eta _0-\eta } \right)+39.74663\left( {\eta _0-\eta }
\right)^2-109.62638\left( {\eta _0-\eta } \right)^3} \over {\eta _0-\eta }}
\label{eq:W2-2}
\end{equation}

\begin{equation}
W={{1-10.589574\left( {\eta _0-\eta } \right)^{2.543}} \over {\left[ {0.694\left( {\eta _0-\eta }
\right)}
\right]^{1.072}}}
\label{eq:W-2}
\end{equation}

\noindent In the preceding expressions $\eta_{0}$ is the packing fraction corresponding to the close
packing density $\rho_{0}$ for the fcc solid.

Finally, parameter $A$ in expression (\ref{eq:g0(x)}) from the condition that the virial theorem 

\begin{equation}
Z_0=1+4\eta g_0\left( 1 \right)
\label{eq:Z-VT}
\end{equation}

\noindent must be satisfied.

\section{RESULTS AND DISCUSSION}

In Figure \ref{fig.1}, the results for the excess energy from the BH and MC perturbation
theories are compared with each other and with the MC simulation data listed in Tables
\ref{Table:MC-gamma=6}-\ref{Table:MC-gamma=36}, and a similar comparison is performed in Figure
\ref{fig.2} for the compressibility factor. The upper limits in the integrals involved
in the BH perturbation theory were set at $x=3.0$, thus making possible a direct comparison with
simulation data without need of including in the latter the corrections due to the truncation of the
potential. Let us first examine the MC simulation data.
At densities $\rho^{*}\lesssim 0.95$ all the systems studied underwent a melting transition. Moreover,
at the lowest temperatures considered the transition showed a tendency, clearly seen in the figures,
to displace towards higher densities ($\rho^{*}\approx 1.0$), and the effect is more pronounced  the
higher is the value of $\gamma$. This is a consequence of the clustering of the particles of these
systems at low temperatures.

Regarding the MC-P results for the excess energy  $U^{E}$, one can see in Figure
\ref{fig.1} that this 'exact' first-order perturbation theory reproduces very accurately the
simulation data for all the values of $\gamma$, at all temperatures and densities except near
melting at low temperatures. The same is true for the compressibility factor $Z$, as shown in Figure
\ref{fig.2}, except for $\gamma\geq 12$ and
$T^{*}=0.6$, the lowest temperature considered, and perhaps sometimes for $T^{*}=0.8$ too. The reason
for the better performance of the MC-P results for the excess energy than for the compressibility
factor is, at least partially, because this 'exact' perturbation theory is second order in the former
quantity whereas it is only first order in the radial distribution function and the equation of state.

Now, let us analyze the performance of the BH perturbation theory. For the excess energy, at
temperatures $T^{*}\geq 1.5$, the contribution of the second-order term is nearly negligible, so that
the BH theory truncated at first order provides satisfactory results in all cases for $T^{*}\geq 1.5$
and even for lower temperatures in the case $\gamma=6$. For $\gamma=12$ and $T\le 1.0$, second-order
approximation is needed, with the local compressibility approximation providing grater accuracy than
the macroscopic compressibility approximation. For $\gamma \geq 18$ and $T\le 1.0$, the second-order
approximation fails to provide completely satisfactory results for the low-density solid. It is to be
remarked that even in the cases where the BH theory departs from simulation data, the 'exact'
first-order MC-P continues doing a good job.

As far as the equation of state is concerned, Figure \ref{fig.2} shows that the BH
perturbation theory is in complete agreement with the MC-P theory up to densities close to the
melting density, except perhaps for $\gamma \le 18$ and $T^{*}=0.6$, with little or no difference
between first- and second-order approximations, although the local compressibility approximation
seems to work slightly better at low temperatures than the first-order or the macroscopic
compressibility approximations.

From the precedent analysis it seems clear that if we were able of improving the theoretical
predictions for the excess energy until reaching the accuracy of the MC-P results, and using the
Helmholtz free energy route to obtain the equation of state instead of the pressure equation, the
results obtained from the perturbation theory would probably be in complete agreement with simulation
data for solids with spherically symmetrical potentials even for extremely short-ranged potentials and
low temperatures. While this improvement is achieved, the Barker-Henderson perturbation theory
constitutes an excellent choice in most situations.

\section*{ACKNOWLEDGMENT}

Financial support by the Spanish Ministerio de Ciencia y Tecnolog\'{\i}a (MCYT) under Grant 
No. BFM2003-001903 is acknowledged.\\

\clearpage

\begin{center}
\Large\bf{List of Tables}
\end{center}

\begin{table}[h!]
\caption{Simulation data for $\gamma= 6$. Corrections due to the truncation of the 
potential have not been included. The number between parenthesis indicates the error in the last
significant digit of the compressibility factor. For the excess energy, the statistical error in
most cases is beyond the third decimal place and when this is the case it is not indicated.}
\begin{center}
\begin{tabular}{lccccccccc}
\hline

$\rho^{*}$ & 0.90 & 0.95 & 1.00 & 1.05 & 1.10 & 1.15  & 1.20 & 1.25 & 1.30 \\
\hline \\

$T^{*}=$0.6  \\
\cline{1-1}

$Z$&  2.70(2) &  3.75(3) &  0.53(2) &  0.28(2) &  0.70(2) &  1.66(2)&  3.84(2) &  8.05(3) &
17.41(4)\\
$U^{E}/N\epsilon$ & -3.382 & -3.651 & -3.852 & -4.149 & -4.479 & -4.832 & -5.208
& -5.610 & -6.036\vspace{0.3 cm}\\

$T^{*}=$0.8  \\
\cline{1-1}

$Z$&  4.59(2) &  5.96(3) &  2.84(2) &  3.04(2) &  3.75(2) &  5.22(2) &  7.71(2) & 12.42(3) &
22.24(4)\\
$U^{E}/N\epsilon$& -3.351 & -3.630 & -3.835 & -4.141 & -4.473 & -4.829 & -5.207 &
-5.609 & -6.036\vspace{0.3 cm}\\

$T^{*}=$1.0  \\
\cline{1-1}

$Z$&  5.85(2) &  7.22(2) &  4.32(2) &  4.72(2) &  5.65(2) &  7.29(2) & 10.05(2) & 15.01(3) &
25.04(4)\\
$U^{E}/N\epsilon$& -3.336 & -3.616 & -3.826 & -4.136 & -4.470 & -4.827 & -5.206 &
-5.609 & -6.036\vspace{0.3 cm}\\

$T^{*}=$1.5  \\
\cline{1-1}

$Z$&  7.481(2) &  9.04(2) &  6.23(2) &  6.95(2) &  8.16(2) & 10.07(2) & 13.15(2) & 18.49(2) &
28.91(4)\\
$U^{E}/N\epsilon$& -3.319 & -3.602 & -3.815 & -4.129 & -4.466 & -4.825 & -5.205 &
-5.609 & -6.036\vspace{0.3 cm}\\

$T^{*}=$2.0  \\
\cline{1-1}

$Z$&  8.277(2) &  9.77(6) &  7.24(2) &  8.06(2) &  9.41(2) & 11.50(2) & 14.73(2) & 20.21(2) &
30.79(4)\\
$U^{E}/N\epsilon$& -3.310& -3.593(1) & -3.810& -4.126& -4.464& -4.824& -5.204& -5.609&
-6.036\vspace{0.3 cm}\\

$T^{*}=$3.0  \\
\cline{1-1}

$Z$&  9.12(2) &  8.18(4) &  8.23(1) &  9.21(2) & 10.74(2) & 12.92(2) & 16.29(2) & 21.92(2) &
32.71(4)\\
$U^{E}/N\epsilon$& -3.303& -3.529(1) & -3.805& -4.123& -4.463& -4.823& -5.204& -5.608&
-6.036\vspace{0.3 cm}\\
\hline

\end{tabular}
\end{center}
\label{Table:MC-gamma=6}
\end{table}

\begin{table}[h!]
\caption{As in Table \ref{Table:MC-gamma=6} for $\gamma= 12$.}
\begin{center}
\begin{tabular}{lccccccccc}
\hline

$\rho^{*}$ & 0.90 & 0.95 & 1.00 & 1.05 & 1.10 & 1.15  & 1.20 & 1.25 & 1.30 \\
\hline \\

$T^{*}=$0.6  \\
\cline{1-1}

$Z$&  4.54(3) &  5.61(3) &  5.98(9) &  2.52(2) &  2.03(3) &  1.86(3) &  2.36(3) &  4.43(3) &
10.884(4)\\
$U^{E}/N\epsilon$& -2.018(1) & -2.221(1) & -2.432(2) & -2.510& -2.760& -3.068& -3.442& -3.892&
-4.429\vspace{0.3 cm}\\

$T^{*}=$0.8  \\
\cline{1-1}

$Z$&  6.06(2) &  7.44(3) &  5.03(3) &  4.61(2) &  4.68(2) &  5.21(2) &  6.54(3) &  9.64(3) &
17.28(4)\\
$U^{E}/N\epsilon$& -1.951(1)& -2.167(1)& -2.247(1)& -2.458& -2.724& -3.046& -3.430& -3.886&
-4.428\vspace{0.3 cm}\\

$T^{*}=$1.0  \\
\cline{1-1}

$Z$&  7.04(2) &  8.36(3) &  5.87(2) &  5.97(2) &  6.40(2) &  7.31(2) &  9.13(3) & 12.77(3) &
21.195(4)\\
$U^{E}/N\epsilon$& -1.919(1) & -2.130(1) & -2.205(1) & -2.431& -2.707& -3.035& -3.423& -3.883&
-4.427\vspace{0.3 cm}\\

$T^{*}=$1.5  \\
\cline{1-1}

$Z$&  8.26(2) &  9.88(2) &  7.23(2) &  7.66(2) &  8.61(2) & 10.09(2) & 12.49(2) & 16.96(3) &
26.23(4)\\
$U^{E}/N\epsilon$& -1.873(1) & -2.089& -2.159& -2.396& -2.683& -3.019& -3.414& -3.880&
-4.426\vspace{0.3 cm}\\

$T^{*}=$2.0  \\
\cline{1-1}

$Z$&  8.86(2) & 10.59(2) &  7.96(2) &  8.60(2) &  9.72(2) & 11.49(2) & 14.17(2) & 19.10(3) &
28.76(4)\\
$U^{E}/N\epsilon$& -1.851& -2.070& -2.139& -2.381& -2.671& -3.011& -3.409& -3.878& -4.426\vspace{0.3
cm}\\

$T^{*}=$3.0  \\
\cline{1-1}

$Z$&  9.46(2) & 10.5(1) &  8.71(2) &  9.53(2) & 10.95(2) & 12.89(2) & 15.93(2) & 21.18(3) &
31.36(4)\\
$U^{E}/N\epsilon$& -1.829& -2.014& -2.119& -2.366& -2.661& -3.004& -3.406& -3.876& -4.425\vspace{0.3 cm}\\
\hline
\end{tabular}
\end{center}
\label{Table:MC-gamma=12}
\end{table}

\begin{table}[h!]
\caption{As in Table \ref{Table:MC-gamma=6} for $\gamma= 18$.}
\begin{center}
\begin{tabular}{lccccccccc}
\hline

$\rho^{*}$ & 0.90 & 0.95 & 1.00 & 1.05 & 1.10 & 1.15  & 1.20 & 1.25 & 1.30 \\
\hline \\

$T^{*}=$0.6  \\
\cline{1-1}

$Z$&  5.411(3) &  6.40(3) &  6.83(7) &  4.17(3) &  3.48(4) &  3.02(3) &  2.68(3) &  2.89(4) & 
6.48(5)\\
$U^{E}/N\epsilon$& -1.604(1) & -1.778(2) & -1.945(3) & -1.995(2) & -2.191(1) & -2.454(1) & -2.790&
-3.231& -3.822\vspace{0.3 cm}\\

$T^{*}=$0.8  \\
\cline{1-1}

$Z$&  6.73(2) &  8.04(3) &  6.81(3) &  5.80(2) &  5.77(3) &  5.95(3) &  6.65(3) &  8.43(4) &
13.85(6)\\
$U^{E}/N\epsilon$& -1.511(1) & -1.695(1) & -1.778(2) & -1.898(1) & -2.116(1) & -2.396& -2.754& -3.214&
-3.817\vspace{0.3 cm}\\

$T^{*}=$1.0  \\
\cline{1-1}

$Z$&  7.58(2) &  8.96(2) &  6.81(3) &  6.88(2) &  7.20(3) &  7.85(3) &  9.09(3) & 11.74(4) &
18.23(4) \\
$U^{E}/N\epsilon$& -1.465(1)& -1.647(1)& -1.673(1) & -1.849(1) & -2.078& -2.368& -2.734& -3.204&
-3.813\vspace{0.3 cm}\\

$T^{*}=$1.5  \\
\cline{1-1}

$Z$&  8.63(2) & 10.25(2) &  7.96(2) &  8.33(2) &  9.15(2) & 10.42(2) & 12.45(3) & 16.23(3) &
24.36(4)\\
$U^{E}/N\epsilon$& -1.402(1) & -1.587(1) & -1.603(1) & -1.787& -2.029& -2.331& -2.710& -3.191&
-3.809\vspace{0.3 cm}\\

$T^{*}=$2.0  \\
\cline{1-1}

$Z$&  9.22(2) & 10.82(2) &  8.46(2) &  9.11(2) & 10.20(2) & 11.71(2) & 14.16(2) & 18.51(3) &
27.37(3)\\
$U^{E}/N\epsilon$& -1.376(1) & -1.558& -1.565(1) & -1.759& -2.006& -2.313& -2.698& -3.186&
-3.807\vspace{0.3 cm}\\

$T^{*}=$3.0  \\
\cline{1-1}

$Z$&  9.73(2) & 11.58(2) &  9.03(2) &  9.86(2) & 11.20(2) & 13.09(2) & 15.89(2) & 20.77(3) &
30.41(3)\\
$U^{E}/N\epsilon$& -1.347(1) & -1.536& -1.533& -1.732& -1.984& -2.297& -2.687& -3.180&
-3.805\vspace{0.3 cm}\\
\hline

\end{tabular}
\end{center}
\label{Table:MC-gamma=18}
\end{table}

\begin{table}[h!]
\caption{As in Table \ref{Table:MC-gamma=6} for $\gamma= 36$.}
\begin{center}
\begin{tabular}{lccccccccc}
\hline

$\rho^{*}$ & 0.90 & 0.95 & 1.00 & 1.05 & 1.10 & 1.15  & 1.20 & 1.25 & 1.30 \\
\hline \\

$T^{*}=$0.6  \\
\cline{1-1}

$Z$&  6.17(2) &  7.20(2) &  7.24(6) &  5.95(3) &  5.86(3) &  5.79(3) &  5.47(3) &  4.88(4) & 
4.11(6)\\
$U^{E}/N\epsilon$& -1.061(2)& -1.194(1) & -1.284(5) & -1.319(2) & -1.460(2) & -1.648(2) & -1.890(1) &
-2.233(1) & -2.753\vspace{0.3 cm}\\

$T^{*}=$0.8  \\
\cline{1-1}

$Z$&  7.58(2) &  8.89(3) &  7.72(4) &  7.61(3) &  8.00(3) &  8.47(3) &  9.03(3) &  9.91(5) &
12.05(6)\\
$U^{E}/N\epsilon$& -0.960(1) & -1.097(1) & -1.100(2) & -1.192(2) & -1.341(1) & -1.536(1) & -1.794(1) &
-2.160& -2.716\vspace{0.3 cm}\\

$T^{*}=$1.0  \\
\cline{1-1}

$Z$&  8.42(2) &  9.80(2) &  8.11(2) &  8.53(2) &  9.07(3) &  9.95(3) & 11.07(3) & 12.92(4) &
16.78(6)\\
$U^{E}/N\epsilon$& -0.907(1) & -1.033(1) & -1.012(1) & -1.126(1) & -1.272(1) & -1.471(1) & -1.739(1) &
-2.119& -2.695\vspace{0.3 cm}\\

$T^{*}=$1.5  \\
\cline{1-1}

$Z$&  9.32(2) & 11.02(3) &  8.81(2) &  9.49(2) & 10.54(2) & 11.90(2) & 13.89(3) & 16.97(4) &
23.19(4)\\
$U^{E}/N\epsilon$& -0.838(1) & -0.970(1) & -0.919(1) & -1.034(1) & -1.188& -1.393& -1.673& -2.070&
-2.669\vspace{0.3 cm}\\

$T^{*}=$2.0  \\
\cline{1-1}

$Z$&  9.65(2) & 11.45(2) &  9.22(2) & 10.01(2) & 11.21(2) & 12.85(3) & 15.28(2) & 19.09(3) &
26.46(4)\\
$U^{E}/N\epsilon$& -0.802& -0.930(1) & -0.877(1) & -0.992& -1.149& -1.357& -1.642& -2.047&
-2.657\vspace{0.3 cm}\\

$T^{*}=$3.0  \\
\cline{1-1}

$Z$& 10.03(2) & 11.59(6) &  9.56(2) & 10.51(2) & 11.89(2) & 13.82(2) & 16.66(2) & 21.19(3) &
29.87(4)\\
$U^{E}/N\epsilon$& -0.771& -0.883(3) & -0.838& -0.954& -1.112& -1.321& -1.611& -2.026&
-2.647\vspace{0.3 cm}\\

\hline
\end{tabular}
\end{center}
\label{Table:MC-gamma=36}
\end{table}

\clearpage

\begin{center}
\Large\bf{List of Figures}
\end{center}

\begin{figure}[h!]
\caption{Comparison between MC simulations (filled circles) and perturbation theory for the
excess energy $U^{E}/N\epsilon$ of Sutherland solids for $T^{*}=0.6,0.8,1.0,1.5,2.0$, and 3.0 from
the bottom upwards. Open circles are the results from the MC perturbation theory as obtained from eq.
(\ref{eq:UE-MC-P}). MC and MC-P results are indistinguishable at the scale of the figure in most
cases. Dotted curves are the results from the first-order Barker-Henderson perturbation theory.
Dashed and continuous curves are the results from the second-order Barker-Henderson perturbation
theory using the macroscopic and local compressibility approximations, respectively. In most cases,
the three curves are nearly indistinguishable to each other at the scale of the figure. For each
temperature the curves, and the corresponding sets of data, have been displaced upwards for clarity
by 1 unit  with respect to those corresponding to the temperature immediately below.}
\end{figure}

\begin{figure}[h]
\caption{Comparison between MC simulations (filled circles) and perturbation theory for the
compressibility factor $Z=pV/Nk_{B}T$ of Sutherland solids for $T^{*}=0.6,0.8,1.0,1.5,2.0$, and 3.0
from the bottom upwards. Open circles are
the results from the MC perturbation theory as obtained from eq. (\ref{eq:Z-MC-P}). Again, MC and MC-P 
results are nearly indistinguishable to each other. The
curves have the same meaning as in fig. \ref{fig.1}. For each
temperature the curves, and the corresponding sets of data, have been displaced upwards for clarity
by 5 units  with respect to those corresponding to the temperature immediately below.}
\end{figure}

\clearpage

\addtocounter{figure}{-2}

\begin{figure}[h!]
\begin{center}
\includegraphics[width=15 cm]{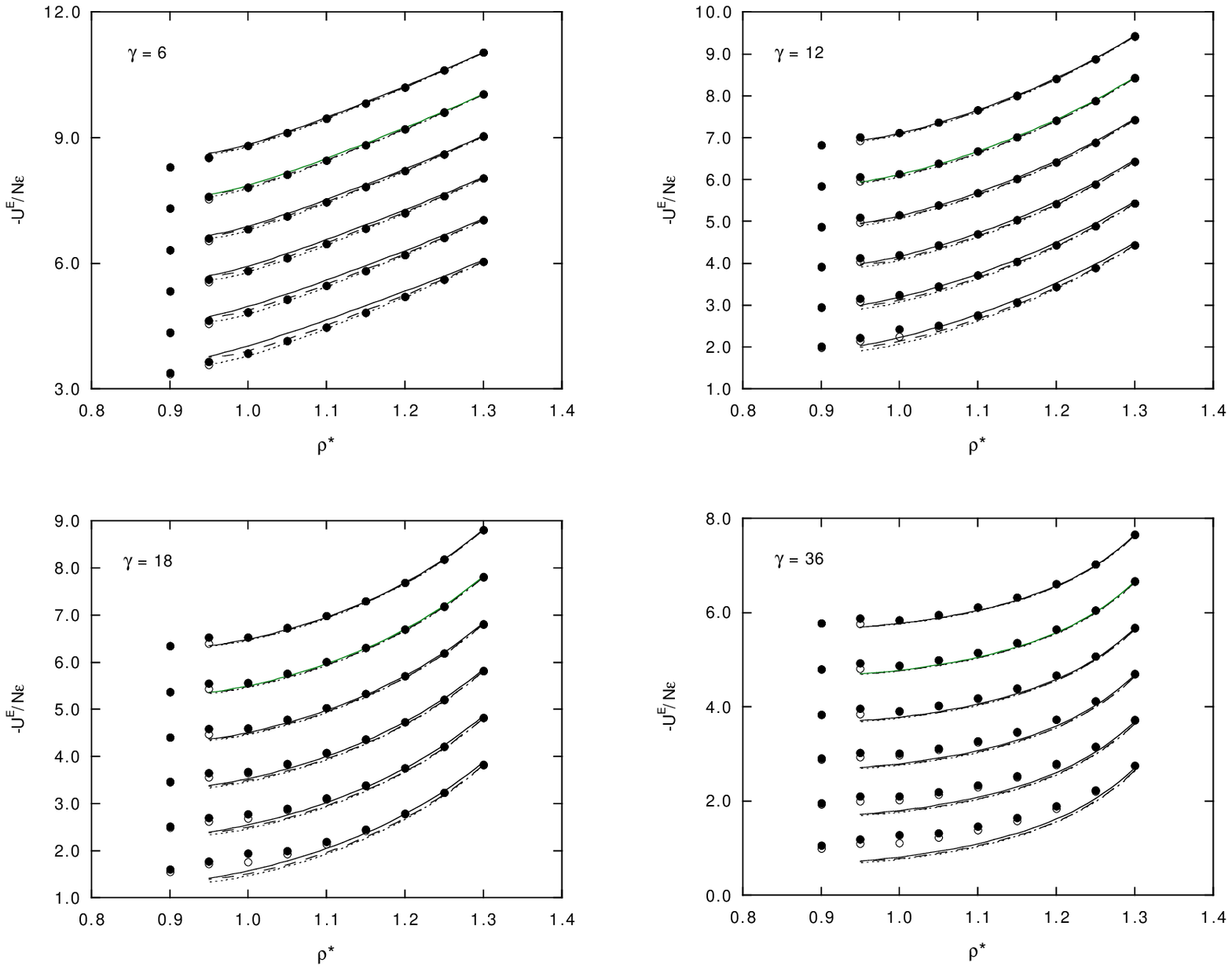}
\caption{}
\label{fig.1}
\end{center}
\end{figure}

\begin{figure}[h!]
\begin{center}
\includegraphics[width=15 cm]{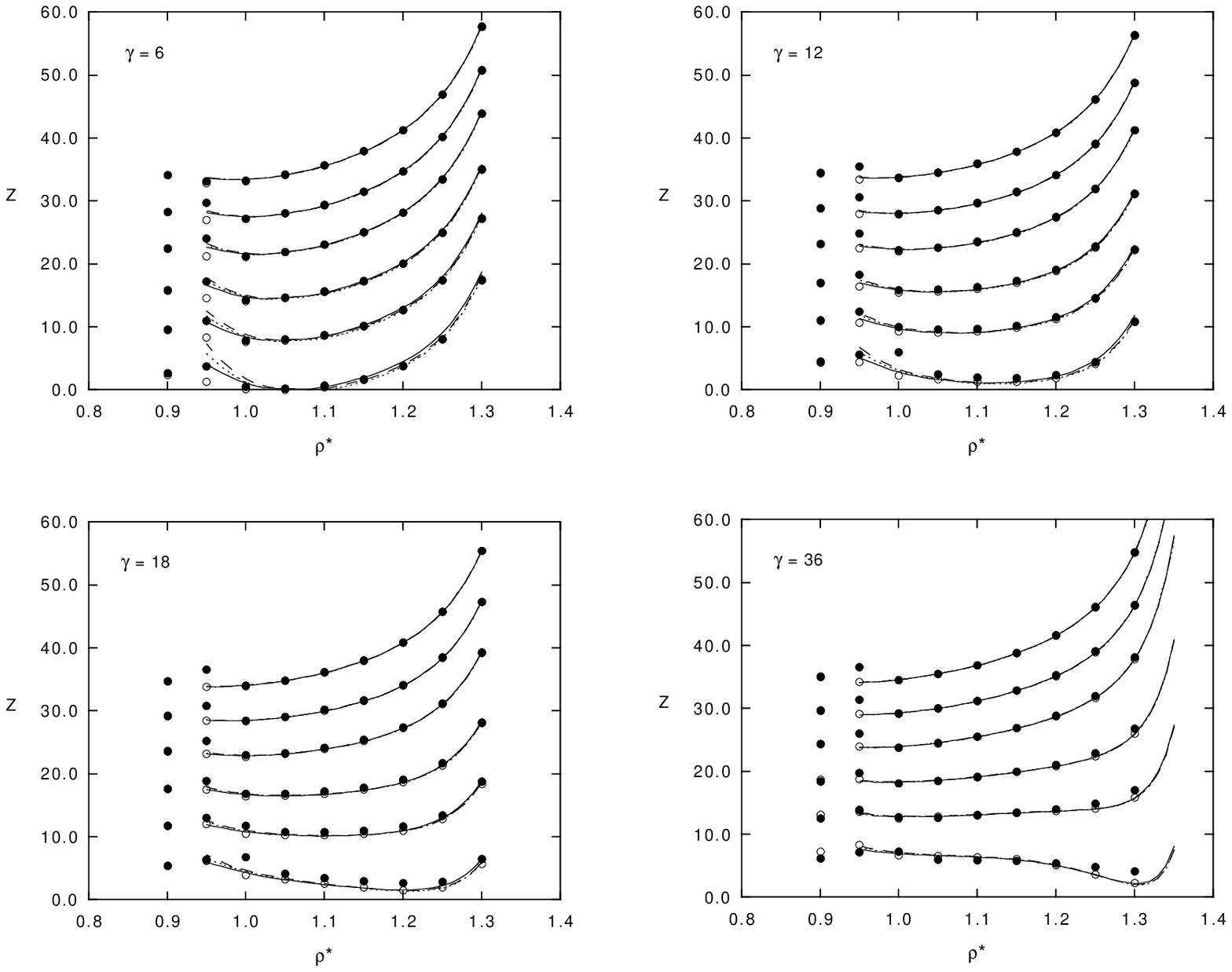}
\caption{}
\label{fig.2}
\end{center}
\end{figure}

\clearpage


\begin{thebibliography}{99}

\bibitem{BS:03} Ben-Amotz, D.; Stell G. {\em J. Chem. Phys.} {\bf 2003}, {\em 119}, 10777-10788.
\bibitem{CVK:00} Coutinho, J. A. P.; Vlamos, P. M.; Kontogeorgis, G. M. {\em Ind. Eng. Chem. Res.}
{\bf 2000}, {\em 39}, 3076-3082.
\bibitem{EP:02} Elvassore, N.; Prausnitz, J. M. {\em Fluid Phase Equil.} {\bf 2002}, {\em 194-197},
567-577.
\bibitem{M:04} Murakami, M. {\em J. Chem. Phys.} {\bf 2004}, {\em 120}, 6751-6755.
\bibitem{ZA:04} Zhang, Q.; Archer, L. A. {\em J. Chem. Phys.} {\bf 2004}, {\em 121}, 10814-10824.
\bibitem{HW:86} Heyes, D. M.; Woodcock, L. V. {\em Mol.  Phys.} {\bf 1986}, {\em 59}, 1369-1388.
\bibitem{DLS:06} D\'{\i}ez, A.; Largo, J.; Solana, J. R. {\em J. of Chem. Phys.} {\bf 2006}, {\em
125},	074509-1-074509-12.
\bibitem{DLS:07} D\'{\i}ez,  A.; Largo, J.; Solana, J. R. {\em Fluid Phase Equil.} In the press.
\bibitem{CP:01} Camp, P. J.; Patey, G. N. {\em J. Chem. Phys.} {\bf 2001}, {\em 114}, 399-408.
\bibitem{C:03} Camp, P. J. {\em Phys. Rev. E} {\bf 2003}, {\em 67}, 011503-1-011503-8.
\bibitem{K:90} Kurochkin, V. I. {\em Sov. Phys. Lebedev Inst. Rep (USA)} {\bf 1990}, {\em 8} 1-2.
\bibitem{LS:00} Largo, J.; Solana, J. R. {\em Int. J. Thermophys.} {\bf 2000}, {\em 21}, 899-908.
\bibitem{J:05} Jiuxun, S. {\em Can. J. Phys.} {\bf 2005}, {\em 83 } 55-66.
\bibitem{BH:67a} Barker, J. A.; Henderson, D. {\em J. Chem. Phys.} {\bf 1967}, {\em 47}, 2856-2861.
\bibitem{BH:72} Barker, J. A.; Henderson, D. {\em Ann. Rev. Phys. Chem.} {\bf 1972}, {\em 23},
439-484.
\bibitem{TL:93} Tang, Y.; Lu, B.C.-Y. {\em J. Chem. Phys.} {\bf 1993}, {\em 99}, 9828-9835.
\bibitem{TL:97} Tang, Y.; Lu, B.C.-Y. {\em Mol. Phys.} {\bf 1997}, {\em 90},	215-224.
\bibitem{SHB:71} Smith, W. R.; Henderson, D.; Barker, J. A. {\em J.  Chem. Phys.} {\bf 1971}, {\em
55}, 4027-4033.
\bibitem{LS:03} Largo, J.; Solana, J. R. {\em Mol. Simulat.} {\bf 2003}, {\em 29}, 363-371.  
\bibitem{LS:03a} Largo, J.; Solana, J. R. {\em Fluid Phase Equil.} {\bf 2003}, {\em 212}, 11-29.
\bibitem{LS:04} Largo, J.; Solana, J. R. {\em J. Phys. Chem. B} {\bf 2004}, {\em 108}, 10062-10070.
\bibitem{BH:67b} Barker, J. A.; Henderson, D. {\em J.  Chem. Phys.}  {\bf 1967}, {\em 47}, 4714-4721.
\bibitem{BH:68} Barker, J. A.; Henderson, D. {\em 4th Symp. Thermophys. Properties} {\bf 1968}, 30-36.
\bibitem{SHB:68} Smith, W. R.; Henderson, D.; Barker, J. A. {\em Can. J. Phys.} {\bf 1968}, {\em
46}, 1725-1727.
\bibitem{SHB:70} Smith, W. R.; Henderson, D.; Barker, J. A. {\em J.  Chem. Phys.} {\bf 1970}, {\em
53}, 508-515.
\bibitem{SHB:75}  Smith, W. R.; Henderson, D.; Barker, J. A. {\em Can. J. Phys.} {\bf 1975}, {\em
53}, 5-12.
\bibitem{HSS:80} Henderson, D.; Scalise, O. H.; Smith, W. R. {\em J. Chem. Phys.} {\bf 1980}, {\em
72}, 2431-2438.
\bibitem{TS:97} Tavares, F. W.; Sandler, S. I. {\em AIChE J.} {\bf 1997}, {\em 43}, 219-231.
\bibitem{TP:04} Tavares, F. W.; Prausnitz, J. M. {\em Colloid Polym. Sci.} {\bf 2004}, {\em 282},
620-632.
\bibitem{Z:04} Zhou, S. {\em J. Phys. Chem. B} {\bf 2004}, {\em 108}, 8447-8451.
\bibitem{H:72} Hall, K. R. {\em J. Chem. Phys.} {\bf 1972}, {\em 57}, 2252-2254.
\bibitem{W:74} Weis, J.-J. {\em Mol. Phys.} {\bf 1974}, {\em 28}, 187-195. Ibid. {\bf 1976}, {\em
32}, 296.
\bibitem{KW:77} Kincaid, J. M.; Weis, J. J. {\em Mol. Phys.} {\bf 1977}, {\em 34}, 931-938.
\bibitem{RMN:96} Rasc\'{o}n, C.; Mederos, L.; Navascu\'{e}s, G. {\em J. Chem. Phys.} {\bf 1996}, {\em
105}, 10527-10534.
\bibitem{VMN:99} Velasco, E.; Mederos, L.; Navascu\'{e}s, G. {\em Mol. Phys.} {\bf 1999}, {\em
97}, 1273-1277.
\bibitem{KRR:86} Kang, H. S.; Ree, T.; Ree, F. H. {\em J. Chem. Phys.} {\bf 1986}, {\em 84},
4547-4557.


\end{thebibliography}
\end{document}